\documentclass[letter,twocolumn]{jpsj2}

\title{
Electrical Resistivity and Thermal Expansion Measurements of URu$_{2}$Si$_{2}$ 
under Pressure
}

\author{
Gaku \textsc{Motoyama}, Nobuyuki \textsc{Yokoyama}, Akihiko \textsc{Sumiyama}, 
and Yasukage \textsc{Oda} 
}

\inst{
Graduate School of Material Science, University of Hyogo, \\
Kamigori-cho, Ako-gun, Hyogo 678-1297, Japan. \\
}

\abst{
We carried out simultaneous measurements of electrical resistivity and thermal expansion 
of the heavy-fermion compound URu$_{2}$Si$_{2}$ under pressure using a single crystal. 
We observed a phase transition anomaly between hidden (HO) and antiferromagnetic (AFM) 
ordered states at $T_{\rm M}$ in the temperature dependence of both measurements. 
For the electrical resistivity, 
the anomaly at $T_{\rm M}$ 
was very small compared with the distinct hump anomaly 
at the phase transition temperature $T_{\rm 0}$ between the paramagnetic state (PM) and HO, 
and exhibited only a slight increase and decrease for the $I$ // $a$-axis and $c$-axis, 
respectively. 
We estimated each excitation gap of HO, $\Delta_{\rm HO}$, and AFM, $\Delta_{\rm AFM}$, 
from the temperature dependence of electrical resistivity; 
$\Delta_{\rm HO}$ and $\Delta_{\rm AFM}$ have different pressure dependences from each other. 
On the other hand, 
the temperature dependence of thermal expansion exhibited a small anomaly at $T_{\rm 0}$ 
and a large anomaly at $T_{\rm M}$. 
The pressure dependence of the phase boundaries of $T_{\rm 0}$ and $T_{\rm M}$ 
indicates that there is no critical end point and the two phase boundaries 
meet at the critical point. 
}

\kword{URu$_{2}$Si$_{2}$, hidden order, antiferromagnetism, 
heavy-fermion superconductor, thermal expansion, electrical resistivity
}

\begin{document}
\maketitle

URu$_2$Si$_2$ is a heavy-fermion superconductor with a 
superconducting transition temperature $T_{\rm c} \sim $ 1.5 K\cite{TTMPal}. 
Furthermore, the compound undergoes a successive phase transition 
at $T_{\rm 0} \sim $ 17.5 K. 
At this temperature, 
specific heat exhibits a sharp and large jump 
of $\sim $ 0.3 J/(K$^{2}\cdot $mol). 
Additionally, there appears a clear kink and a clear hump 
in the curves of magnetic susceptibility 
and electrical resistivity plotted as a function of temperature $T$, respectively\cite{MBMapl}. 
These features show a weak sample dependence. 
On the other hand, in many neutron diffraction experiments, 
only a tiny staggered moment of about $\sim $ 0.03 $\mu _{\rm B}$/U was observed, 
and there were strong sample dependences on its magnitude 
and onset\cite{CBroho, BFaK, THonm, HAmi1}. 
These results have led to many speculations that the true order parameter is not 
the weak magnetic dipole moment, 
but another unknown symmetry such as quadrupoles.

Amitsuka {\it et al.} presented neutron diffraction data 
obtained under high pressure\cite{HAmi2}. 
They observed that the staggered moment increased with increasing pressure $P$ 
and also pointed out that the 3D Ising type of antiferromagnetic phase (AFM) exists 
above the critical pressure $P_{\rm c} \sim $ 15 kbar. 
Since their study, some measurements to study the AFM phase under high pressure have 
been carried out\cite{HAmi1, KMatsu, GMoto1, GMoto2, AAmato, MYoko, SUemu,NKSato, FBourd, EHass}. 
The high-pressure $^{29}$Si-NMR measurements by Matsuda {\it et al.} indicated that the 
AFM volume fraction develops spatially inhomogeneously upon pressure application\cite{KMatsu}. 
One of authors of this Letter and collaborators 
performed thermal expansion measurements under pressure to obtain thermodynamical 
evidence of the presence of the $P$-induced AFM ordering\cite{GMoto1, GMoto2}. 
A phase transition between the hidden ordered state (HO) and AFM at $T_{\rm M}$ was found; 
the $P$ dependence of $T_{\rm M}$ ($T_{\rm M}$($P$)) was revealed. 
The authors suggested that the first-order-like $T_{\rm M}$($P$) 
and the second-order $T_{\rm 0}$($P$) meet at $P_{\rm c}$, 
and second-order $T_{\rm N}$($P$) exists above $P_{\rm c}$. 
Uemura {\it et al.} examined the $T$ dependence of dc magnetization under pressure, 
and observed the anomaly at the phase transition from HO to AFM\cite{SUemu, NKSato}. 
The authors argued the presence of the bicritical point on the basis of the $P$ dependence of the 
parasitic ferromagnetic anomaly $T_{\rm FM}$($P$) of $\sim$ 35 K for $P$ = 0. 
Moreover, 
it was revealed that the superconductivity of this system coexists only in HO but not in AFM. 
However, Bourdarot {\it et al.} argued the presence of the critical end point 
of the $T_{\rm M}$($P$) on the basis of their neutron diffraction measurements\cite{FBourd}. 
Recently, Hassinger {\it et al.} observed the anomaly at the phase transition from HO to AFM 
in the electrical resistivity and specific heat, 
and showed the $P$ dependences of these phase boundaries 
that met at the critical point\cite{EHass}. 
Whether or not $T_{\rm M}$($P$) and $T_{\rm 0}$($P$) meet is important information 
concerning the symmetry of the order parameter of the HO state\cite{VPMine}. 
However past experiments are insufficient to conclude 
whether or not $T_{\rm M}$($P$) and $T_{0}$($P$) meet. 
The thermal expansion measurement was sensitive to $T_{\rm M}$ but it yielded no details 
of $T_{0}$, particularly at around $P_{\rm c}$. 
A smaller anomaly was smeared out within a predominant anomaly 
when $T_{\rm M}$ approached $T_{0}$. 
On the other hand, electrical resistivity and specific heat were sensitive to only $T_{0}$, 
that is, these measurements yielded no details of $T_{\rm M}$ at around $P_{\rm c}$. 
In this work, 
we carried out electrical resistivity and thermal expansion measurements in parallel. 
In the entire $P$ range, even at around $P_{\rm c}$, 
$T_{\rm M}$ and $T_{0}$ could be accurately determined from the data of 
thermal expansion and electrical resistivity, respectively, 
which were measured at the same time under pressure to avoid the ambiguity 
between the two measurements.

We first synthesized a polycrystalline material by melting a stoichiometric amount 
of the constituent elements natural U, Ru, and Si, 
which had purities of 99.9 \%, 99.99 \%, and 99.9999 \%. 
Then we grew a single crystal by the Czochralski pulling method 
from the polycrystalline material in high-purity argon atmosphere 
using a laboratory-made tri-arc furnace. 
The sample for measurements was cut from the as-grown single crystal. 
The size of the sample was about $\sim $ 2$\times $2$\times $2 mm$^{3}$. 
We chose a sample that showed a distinct anomaly at $T_{\rm M}$ 
in order to determine $T_{0}$($P$) and $T_{\rm M}$($P$), 
because the anomaly of $T_{\rm M}$ exhibits a strong sample dependence, 
whereas the anomaly of $T_{0}$ has only a weak dependence. 
We measured electrical resistivity and thermal expansion 
by the conventional dc 4-terminal method and 
the strain gauge technique with a copper block as a dummy sample, respectively. 
The measurements were performed using a $^{4}$He cryostat down to 4 K. 
Pressure was generated using a copper-beryllium clamp-type cylinder 
with a piston made of tungsten carbide. 
The pressure-transmitting medium was Dafune7373. 
We determined pressure by measuring the superconducting 
transition temperature of indium. 
The electrical resistivity and thermal expansion measurements were carried out 
concurrently to eliminate measurement ambiguities in pressure and temperature.

\begin{figure}
\begin{center}
\includegraphics[width=8.5cm]{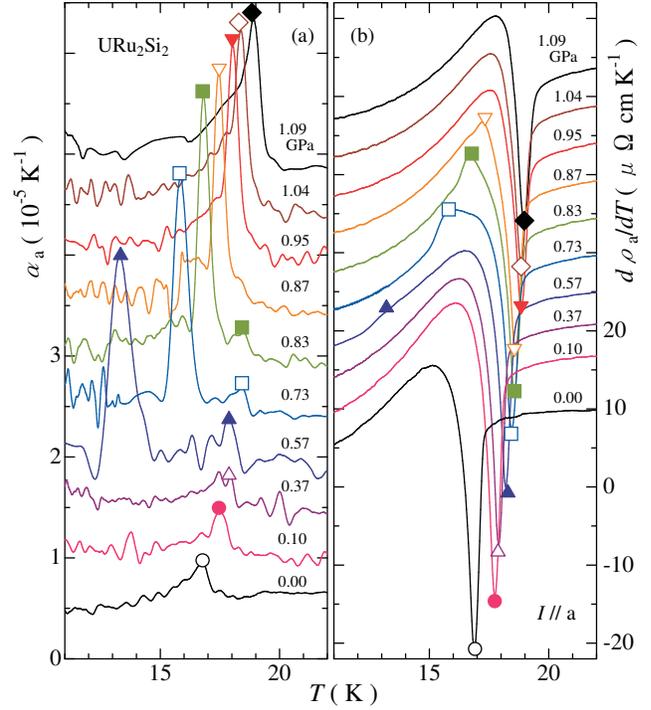}
\end{center}
\caption{ (Color online) 
(a) $T$ dependence of the thermal expansion coefficient for the $a$-axis, $\alpha_{\rm a}$, 
at different pressures. 
(b) $T$ dependence of the derivative of the resistivity 
for the $a$-axis, $d\rho_{\rm a}/dT$, at different pressures. 
The curves are shifted along the vertical axis for clarity. 
$\alpha_{\rm a}$($T$) and $d\rho_{\rm a}/dT$($T$) 
data were measured concurrently, namely, 
there is no ambiguity in $P$ and $T$ between $\alpha_{\rm a}$ and $d\rho_{\rm a}/dT$. 
The marks indicate the positions of the maximum or minimum of the anomalies 
at $T_{0}$ and $T_{\rm M}$ at the different pressures 
(from bottom to top: $P$ = 0.00 ($\circ$), 0.10 ($\bullet$), 0.37 ($\vartriangle$), 
0.57 ($\blacktriangle$), 0.73 ($\square$), 0.83 ($\blacksquare$), 0.87 ($\triangledown$), 
0.95 ($\blacktriangledown$), 1.04 ($\lozenge$), and 1.09 GPa ($\blacklozenge$)). 
An anomaly at $T_{0}$ was observed at $\sim$ 17 K and 0 GPa 
in both measurements. 
At 0.57 GPa and 13.5 K, there appears another anomaly of $T_{\rm M}$ in both measurements. 
Note that $\alpha_{\rm a}$ is sensitive to the transition at $T_{\rm M}$, 
whereas $d\rho_{\rm a}/dT$ is sensitive to the transition at $T_{0}$. 
It is easy to determine $T_{\rm M}$($P$) and $T_{0}$($P$) by examining 
$\alpha_{\rm a}$($T$) and $d\rho_{\rm a}/dT$($T$), respectively. 
} 
\label{f1}
\end{figure}

Figure 1(a) shows the $T$ dependence of the thermal expansion coefficient 
for the $a$-axis, $\alpha_{\rm a}$, at different pressures. 
A mean-field-like discontinuous anomaly corresponding to the phase transition at $T_{0}$ 
between the paramagnetic state (PM) and HO was observed at 17 K at 0 GPa, 
and it could be observed only below 0.83 GPa 
within the accuracy of the $\alpha_{\rm a}$ measurement. 
An anomaly that was identified as the phase transition between the HO and AFM 
appeared at 0.57 GPa and $T_{\rm M}$ $\sim $ 13.5 K. 
This anomaly was greater than the anomaly at $T_{0}$. 
It is clear that $T_{\rm M}$ shifts to higher temperatures accompanied by a change 
in the shape of the anomaly, and finally becomes a large mean-field-like anomaly. 
These behaviors are the same as those described in our previous paper\cite{GMoto1}. 
Next, Fig. 1(b) shows the $T$ dependence of the derivative of the resistivity 
for the $a$-axis, $d\rho_{\rm a}/dT$, at different pressures. 
There is a sharp and deep dip at $T_{0}$ in $d\rho_{\rm a}/dT$($T$). 
It shifts to higher temperatures, remaining sharp and deep, 
throughout the entire range of pressure. 
On the other hand, a small convex-upward anomaly is also seen at $T_{\rm M}$ 
in $d\rho_{\rm a}/dT$($T$) above 0.57 GPa. 
This anomaly was too small, in comparison with the deep dip, 
to observe the $P$ dependence of $T_{\rm M}$. 
When $T_{\rm M}$ was close to $T_{0}$, it was smeared out in the dip. 
The measurements of $\alpha_{\rm a}$ and $d\rho_{\rm a}/dT$ were carried out at the same time. 
Therefore, the $P$ and $T$ of $\alpha_{\rm a}$ are identical to those of $d\rho_{\rm a}/dT$, 
although there may be a slight error in absolute value. 
$T_{0}$ and $T_{\rm M}$ were defined 
as the maximum temperature in the data of both $\alpha_{\rm a}$ and $d\rho_{\rm a}/dT$, 
and are indicated by marks in Figs. 1(a) and 1(b). 
The error ranges of $T_{0}$ and $T_{\rm M}$ were determined from the full width at half-maximum 
of the peak or the full width at half-minimum of the dip. 
The $P$ dependences of $T_{0}$ and $T_{\rm M}$ are plotted in Fig. 4(a). 
These results are described below.

\begin{figure}
\begin{center}
\includegraphics[width=8.5cm]{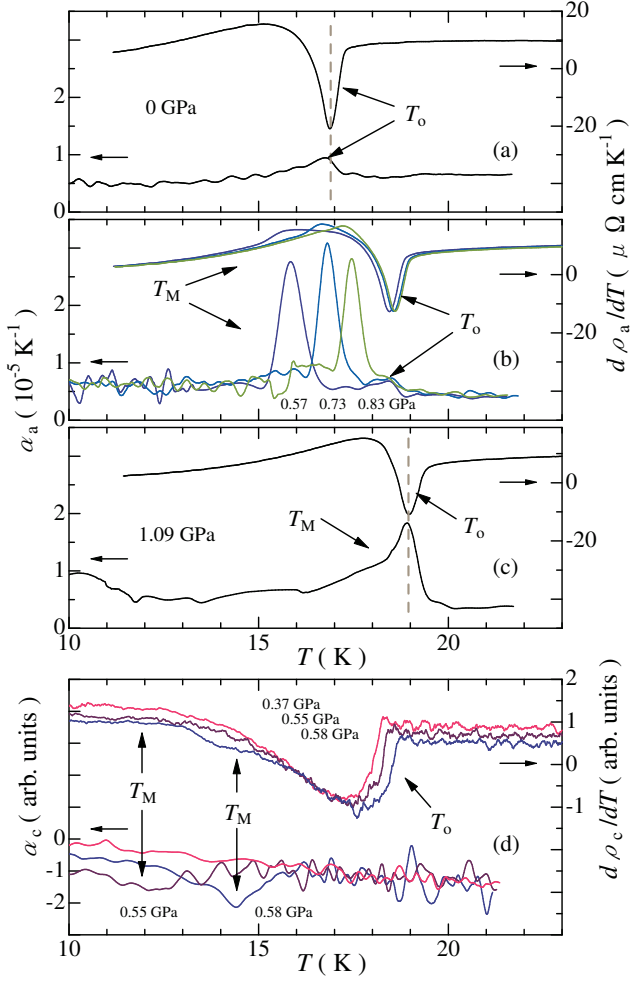}
\end{center}
\caption{ (Color online) 
(a) $T$ dependences of $\alpha_{\rm a}$ and $d\rho_{\rm a}/dT$ at 0 GPa. 
Only the anomaly was observed at $T_{0}$. 
(b) The same as in (a) but for 0.57, 0.73, and 0.83 GPa. 
At these pressures, the anomalies at $T_{0}$ and $T_{\rm M}$ are observed in both 
measurements. 
(c) The same as in (a) but for 1.09 GPa. 
At this pressure, the anomaly at $T_{0}$ was observed in $d\rho_{\rm a}/dT$ data, 
whereas the anomaly at $T_{\rm M}$ was observed in $\alpha_{\rm a}$ data. 
(d) $T$ dependences of $\alpha_{\rm c}$ and $d\rho_{\rm c}/dT$ 
at 0.37, 0.55, and 0.58 GPa. 
The curves of $d\rho_{\rm c}/dT$ are shifted along the vertical axis for clarity. 
} 
\label{f2}
\end{figure}

Figures 2(a)-2(c) represent the $T$ dependences of $\alpha_{\rm a}$ 
and $d\rho_{\rm a}/dT$ at 0, 0.57, 0.73, 0.83, and 1.09 GPa. 
At ambient pressure, we should observe only the phase transition between PM and HO. 
Our $\alpha_{\rm a}$ and $d\rho_{\rm a}/dT$ data certainly showed the anomaly at the 
same temperature $T_{0}$. 
Next, in Fig. 2(b), we could observe the $P$ dependence of 
the anomalies of $T_{0}$ and $T_{\rm M}$ from 0.57 to 0.83 GPa; 
$\alpha_{\rm a}$ is sensitive to the transition at $T_{\rm M}$, 
while $d\rho_{\rm a}/dT$ is sensitive to the transition at $T_{0}$. 
Moreover, $d\rho_{\rm a}/dT$($T$) 
evidently shows a convex-upward anomaly at $T_{\rm M}$. 
When $d\rho_{\rm a}/dT$($T$) has a convex-upward anomaly, 
$\rho_{\rm a}$($T$) must have a steplike anomaly at $T_{\rm M}$. 
We show $\rho_{\rm a}$($T$) in Fig. 3; 
these results are described below. 
At 1.09 GPa, in Fig. 2(c), we observed a large peak of $\alpha_{\rm a}$ and 
a deep dip of $d\rho_{\rm a}/dT$ at the same temperature. 
The large peak of $\alpha_{\rm a}$ corresponds to $T_{\rm M}$ and 
the deep dip of $d\rho_{\rm a}/dT$ corresponds to $T_{0}$. 
Therefore, $T_{0}$ and $T_{\rm M}$ have the same value at 1.09 GPa. 
The phase boundaries of $T_{0}$ and $T_{\rm M}$ met and constructed the phase transition 
between PM and AFM at the temperature $T_{\rm N}$. 
Moreover, the anomaly of the large peak of $\alpha_{\rm a}$ 
was retained upto 1.69 GPa in our previous study\cite{GMoto1}, 
and the anomaly of $\rho$($T$) was retained upto over $\sim $ 2 GPa 
in previous studies\cite{EHass, MWMcE}. 
We consider that the anomalies observed at $T_{0}$ and $T_{\rm M}$ occur 
at the same temperature as the anomaly of $T_{\rm N}$ at pressures higher than $P_{\rm c}$. 
Figure 2(d) shows the $T$ dependences of the thermal expansion coefficient 
for the $c$-axis, $\alpha_{\rm c}$, and the derivative of the resistivity 
for the $c$-axis, $d\rho_{\rm c}/dT$, at 0.37, 0.55, and 0.58 GPa. 
There is a large anisotropy between $\rho_{\rm a}$ and $\rho_{\rm c}$; 
$\rho_{\rm c}$ is one-tenth of $\rho_{\rm a}$. 
Therefore, it was difficult to obtain the absolute value of $\rho_{\rm c}$; 
consequently, we show $d\rho_{\rm c}/dT$ and $\alpha_{\rm c}$ in arbitrary units. 
We also observed anomalies at $T_{\rm M}$ in $\alpha_{\rm c}$ and $d\rho_{\rm c}/dT$, 
although these anomalies were small. 
These small anomalies were consistent with the previous results\cite{GMoto1, EHass}. 
Here, note that the anomaly in $d\rho_{\rm c}/dT$ at $T_{\rm M}$ is 
convex-downward. 
The anomaly in $\rho$ at $T_{\rm M}$ clearly exhibits anisotropy.

\begin{figure}
\begin{center}
\includegraphics[width=8.5cm]{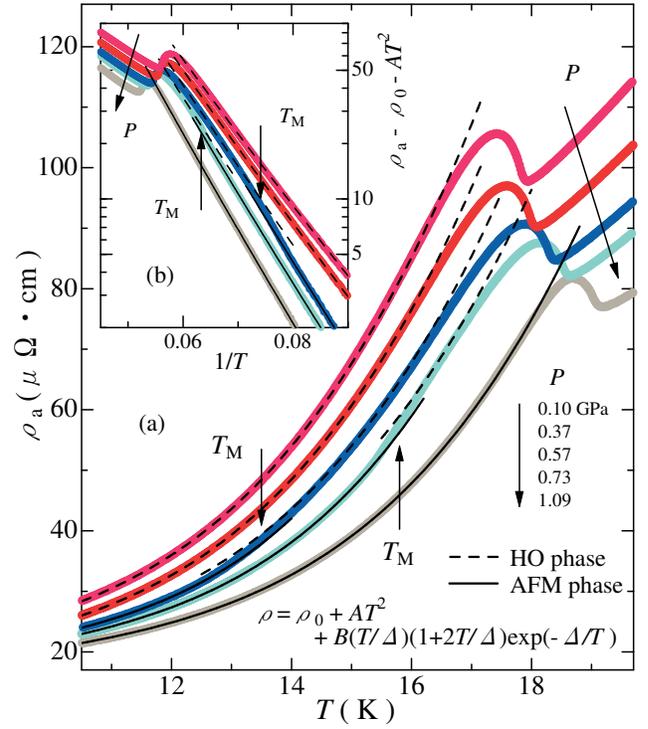}
\end{center}
\caption{ (Color online) 
(a) $T$ dependence of $\rho_{\rm a}$ at 0.10, 0.37, 0.57, 0.73, and 1.09 GPa. 
(b) The inset is the plot of $\rho_{\rm a}$-$\rho_{0}$-$AT^{2}$ on a logarithmic scale 
vs 1/$T$. 
The lines represent the fit of eq. (1) to the data. The full and broken lines 
correspond to the AFM phase and HO phase data. 
} 
\label{f3}
\end{figure}

Figure 3 shows the $T$ dependence of $\rho_{\rm a}$ at 0.10, 0.37, 0.57, 0.73, and 1.09 GPa. 
Previous $\rho$($T$) results for URu$_{2}$Si$_{2}$ could be fitted by the sum of 
the $T^{2}$ term and the exp$(-\Delta /T)$ term\cite{EHass, MWMcE, SAMMen}: 
\begin{eqnarray}
\label{eq:1}
\rho&=&\rho_{0} + AT^2 + B\frac{T}{\Delta}(1+2\frac{T}{\Delta}){\rm exp}(\frac{-\Delta}{T}). 
\end{eqnarray}  
\noindent
We attempted to fit this equation to our $\rho_{\rm a}$ data. 
$\rho_{\rm a}$ was expected to be fitted easier than $\rho_{\rm c}$ for the excitation feature 
because of the strong $T$ dependence of $\rho_{\rm a}$. 
Our $\rho_{\rm a}(T)$ data below 0.37 GPa could be fitted well with eq. (1). 
In this pressure region, the HO phase exists below $T_{0}$. 
That is, the $\rho_{\rm a}(T)$ of the HO region could be fitted with eq. (1). 
Moreover, the $\rho_{\rm a}(T)$ data at 1.09 GPa could also be fitted well. 
At this pressure, the phase boundaries of $T_{0}$($P$) and $T_{\rm M}$($P$) meet; 
therefore, the AFM phase exists below the anomaly at $T_{\rm M}$. 
$\rho_{\rm a}(T)$ in the AFM region could also be fitted with eq. (1) 
using the appropriate parameters for the AFM state. 
However, the $\rho_{\rm a}$($T$) data from 0.57 to 0.87 GPa 
show a steplike anomaly at $T_{\rm M}$. 
When there is a steplike anomaly, we must fit separately at $T_{\rm M}$. 
The lower and higher parts of data were fitted with eq. (1) 
using the appropriate parameters for HO and AFM states, respectively. 
It is difficult to discuss the $T$ dependence of $\rho_{\rm a}$ at around $T_{\rm M}$ 
because of the inevitable phase separation of first-order transition. 
We must estimate these parameters without $\rho_{\rm a}$($T$) data at around $T_{\rm M}$. 
The $P$ dependences of the excitation gaps, $\Delta_{\rm HO}$ and $\Delta_{\rm AFM}$, and 
the coefficients of the $T^2$ contribution, $A_{\rm HO}$ and $A_{\rm AFM}$, 
are plotted in Figs. 4(b) and 4(c), respectively. 
In our estimation, there were small differences in the accuracy 
between $\rho_{0 \rm ,HO}$ and $\rho_{0 \rm ,AFM}$ and 
between $A_{\rm HO}$ and $A_{\rm AFM}$ for $\rho_{\rm a}$($T$) at 0.57 and 0.73 GPa, 
respectively. 
Therefore, we show electrical resistivity data without the residual resistivity and 
Fermi liquid contribution $\rho_{\rm a}$-$\rho_{0}$-$AT^{2}$ on a logarithmic scale 
as a function of 1/$T$ in the inset of Fig. 3. 
The decreasing rates of log($\rho_{\rm a}$-$\rho_{0}$-$AT^{2}$) at 0.10, 0.37, and 1.09 GPa 
are almost constant within the plotted $T$ range, 
although these rates become slightly slow 
owing to the coefficient of the exponential term of $\rho_{\rm a}$. 
The decreasing rate of log($\rho_{\rm a}$-$\rho_{0}$-$AT^{2}$) vs 1/$T$ roughly 
corresponds to $\Delta_{\rm HO}$ or $\Delta_{\rm AFM}$. 
When the AFM phase appears, namely, a broken line turns into a full line, 
the rate becomes more rapid, 
indicating that $\Delta_{\rm AFM}$ $\neq$ $\Delta_{\rm HO}$. 
It is natural to have different excitation gaps 
for different ordered states. 
It is a interesting that eq. (1) well fits not only the $\rho_{\rm a}$($T$) 
of the AFM phase but also that of the HO phase. 
This result may provide a clue to the order parameter of the HO phase.

\begin{figure}
\begin{center}
\includegraphics[width=7.4cm]{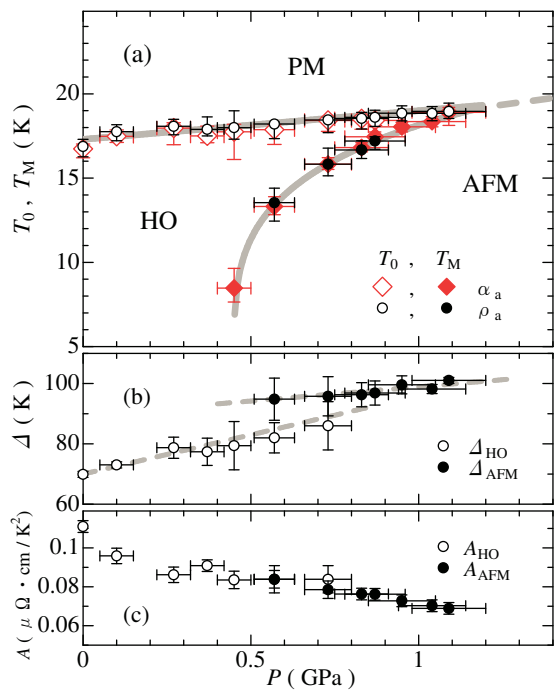}
\end{center}
\caption{ (Color online) 
(a) $P$ dependences of $T_{0}$ ($\circ$, $\lozenge$) and $T_{\rm M}$ ($\bullet$, $\blacklozenge$), 
as derived from maximum and minimum temperatures of 
$d\rho_{\rm a}/dT$($T$) ($\circ$, $\bullet$) and 
$\alpha_{\rm a}$($T$) ($\lozenge$, $\blacklozenge$). 
The lines are guides for the eye. 
(b) $P$ dependences of $\Delta_{\rm HO}$ ($\circ$) and $\Delta_{\rm AFM}$ ($\bullet$), 
as derived from fits of eq. (1) to the $\rho_{\rm a}$ data of the HO and AFM phases, 
respectively. 
The lines are guides for the eye. 
(c) $P$ dependences of the coefficient of the $T^2$ contribution $A_{\rm HO}$ ($\circ$) and 
$A_{\rm AFM}$($\bullet$), derived as in (b). 
} 
\label{f4}
\end{figure}

In Fig. 4(a), we summarize the $P$-$T$ phase diagram of URu$_{2}$Si$_{2}$ using data from 
$\rho_{\rm a}$($T$) and $\alpha_{\rm a}$($T$) measurements; 
it includes details about $T_{0}$ and $T_{\rm M}$ at around $P_{\rm c}$. 
It was experimentally verified that 
$T_{\rm M}$ meets $T_{0}$ at the critical point, 
where $P_{\rm c}$ is from 1.04 to 1.09 GPa in this sample. 
We took measurements only below 1.09 GPa because of the limit of our pressure cell. 
In Figs. 4(b) and 4(c), we plot the $P$ dependences of $\Delta_{\rm HO}$, 
$\Delta_{\rm AFM}$ and $A_{\rm HO}$, $A_{\rm AFM}$, respectively. 
$\Delta_{\rm HO}$($P$) and $\Delta_{\rm AFM}$($P$) have different $P$ dependences from each other. 
However, $\Delta_{\rm HO}$($P$) and $\Delta_{\rm AFM}$($P$) also increase gradually 
with increasing $P$, 
and also seem to show linear $P$ dependences. 
These extrapolated lines seem to cross at around $P_{\rm c}$. 
$A_{\rm HO}$($P$) and $A_{\rm AFM}$($P$) decrease gradually with increasing $P$. 
Although the difference in $P$ dependence between 
$A_{\rm HO}$($P$) and $A_{\rm AFM}$($P$) cannot be denied, 
the differences between $A_{\rm HO}$ and $A_{\rm AFM}$ are negligible 
in terms of the accuracy of the measurements and estimations.

In conclusion, 
our results include two significant points to be emphasized. 
The first one is that the HO and AFM phases are completely separated by the boundary 
of $T_{\rm M}$, which seems to be a first-order transition. 
The second one is that each of the HO and AFM phases also has an excitation gap; 
$\Delta_{\rm HO}$($P$) was not identical to $\Delta_{\rm AFM}$($P$). 
These two results clearly indicate that the HO state is not identical to the AFM state.

\section*{Acknowledgments}
We thank N. K. Sato, T. Kohara, Y. Takahashi and Y. Hasegawa for helpful discussions.

\end{document}